\documentclass{article}

\usepackage{spconf,amsmath,graphicx}

\usepackage[hidelinks, breaklinks]{hyperref}

\usepackage{xcolor}
\usepackage{amssymb}
\usepackage{multirow}
\usepackage{booktabs}
\usepackage{acronym}

\title{Blind estimation of audio effects using an auto-encoder approach and differentiable digital signal processing}
\name{Côme Peladeau and Geoffroy Peeters\thanks{This work was performed under fundings by the 'Hi! PARIS Call for Collaborative and scientific Projects 2022' and the 'AQUA-RIUS project founded by the ANR-22-CE23-0022 program'.}}

\address{LTCI - Télécom-Paris, Institut Polytechnique de Paris, France}

\date{August 2023}

\newcommand{\hatbf}[1]{\ensuremath{\hat{\mathbf{#1}}}}

\newcommand{\norm}[1]{\ensuremath{\left\Vert #1 \right\Vert}}

\DeclareMathOperator{\sgn}{sign}

\newcommand{\x}{\ensuremath{\mathbf{x}}}
\newcommand{\q}{\ensuremath{\mathbf{q}}}
\newcommand{\hq}{\ensuremath{\hatbf{q}}}
\newcommand{\p}{\ensuremath{\mathbf{p}}}
\newcommand{\hp}{\ensuremath{\hatbf{p}}}
\newcommand{\y}{\ensuremath{\mathbf{y}}}
\newcommand{\hy}{\ensuremath{\hatbf{y}}}
\newcommand{\Myy}[0]{\ensuremath{\text{MSE}_{\hy,\y}}}
\newcommand{\Mqq}[0]{\ensuremath{\text{MSE}_{\hq,\q}}}
\newcommand{\Lyy}[0]{\ensuremath{\mathcal{L}^\text{Mel}_{\hy,\y}}}

\acrodef{tcn}[TCN]{Temporal Convolution Network}
\acrodef{film}[FiLM]{Feature-Wise Linear Modulation}
\acrodef{np}[NP]{Neural Proxy}
\acrodef{beafx}[BE-AFX]{Blind Estimation of Audio Effects}
\acrodef{afx}[AFX]{audio effect}

\acrodef{drc}[DRC]{dynamic range compressor}

\newenvironment{customequation}{
	\vspace{-.1cm}
	\begin{equation}
	}{
		\vspace*{-.1cm}
	\end{equation}
}

\makeatletter
\def\blfootnote{\gdef\@thefnmark{}\@footnotetext}
\makeatother

\begin{document}
\ninept

\maketitle
\begin{abstract}
\ac{beafx} aims at estimating the \acp{afx} applied to an original, unprocessed audio sample solely based on the processed audio sample.
To train such a system traditional approaches optimize a loss between ground truth and estimated \ac{afx} parameters. 
This involves knowing the exact implementation of the \acp{afx} used for the process.
In this work, we propose an alternative solution that eliminates the requirement for knowing this implementation. 
Instead, we introduce an auto-encoder approach, which optimizes an audio quality metric. 
We explore, suggest, and compare various implementations of commonly used mastering \acp{afx}, using differential signal processing or neural approximations. 
Our findings demonstrate that our auto-encoder approach yields superior estimates of the audio quality produced by a chain of \acp{afx},  compared to the traditional parameter-based approach, even if the latter provides a more accurate parameter estimation.
\end{abstract}


\vspace{-0.1cm}
\section{Introduction}

\vspace{-0.1cm}
Audio effects (AFXs) play an essential role in music production.
They are used during mixing to sculpt sounds for artistic purposes or context requirements (such as when a sound needs to be mixed with others).
They are used during mastering, the final stage of production, to improve the clarity of a given mix, adapt it for a given media (such as vinyl or streaming), or harmonize it with other tracks in the album.
For these reasons, its automatization has been the subject of several softwares\footnote{
    such as Izotope Ozone 11 or Sonible smart:EQ 4.
}
which allow learning the mastering EQ of a given track to apply it to another.
In this work, we study the generalization to other common mastering \acp{afx}.


\acf{beafx} aims at estimating the \acfp{afx} applied to an original, unprocessed (dry) audio sample \x{} solely based on the observation of the processed (wet) audio sample \y.
This estimation takes the form of the \acp{afx} and their parameters \p.
\vspace{-0.3cm}
\subsection{Related works}
For long \ac{beafx} techniques have been based on explicit rules and assumptions. 
For example, Ávila et al.~\cite{avilaMLEstimationMemoryless2014} proposed to estimate the curve of a memoryless non-linear distortion by assuming that the unprocessed signal has the statistics of a Gaussian white noise.
However, nowadays, most \ac{beafx} approaches rely on training neural networks.
Indeed, following SincNet~\cite{ravanelliSpeakerRecognitionRaw2018} and DDSP~\cite{engelDDSPDifferentiableDigital2020}, modeling audio processes as differential processes has allowed developing differentiable \acp{afx} as specialized neural networks layers with interpretable parameters~\cite{nercessianNeuralParametricEqualizer2020,leeDifferentiableArtificialReverberation2022,steinmetzStyleTransferAudio2022}. 
Because they are differentiable, they can be integrated transparently in a neural network.
Since then, differentiable \acp{afx} have been used for many tasks:  automatic mixing and mastering \cite{steinmetzAutomaticMultitrackMixing2021,martinezramirezDifferentiableSignalProcessing2021}, production style transfer \cite{steinmetzStyleTransferAudio2022} or estimation of audio effects \cite{colonelReverseEngineeringRecording2021}.

In the case of \ac{beafx}, neural networks are usually trained to minimize a loss function that aims at reconstructing the \ac{afx} chain and its parameters as did Hinrichs et al.~\cite{hinrichsClassificationGuitarEffects2022} or Lee et al.~\cite{leeBlindEstimationAudio2023}.
However, while their approach is flexible, its training requires the knowledge of the used \acp{afx} and their parameters. Our approach does not.
Also, as we will highlight in this work, a parameter distance does not translate well to a perceptual distance between audio effects.
This is why we will propose here the use of an audio distance.

Estimation of audio effects with an audio loss function and differentiable audio effects has already been investigated.
For example, Colonel et al.~\cite{colonelReverseEngineeringMemoryless2022} used it for non-linear distortion using a differentiable Wiener-Hammerstein model and also in \cite{colonelReverseEngineeringRecording2021} for a complete mixing setting. 
However, in both cases, their approaches require paired \x{} and \y{} data for the estimation, i.e. they did not perform the blind estimation.
In this work, we perform blind estimation, i.e. we aim at estimating the \acp{afx} applied to $\x$ using only the knowledge of $\y$.

\begin{figure*}[ht]
    \centering
\includegraphics[width=.9\textwidth]{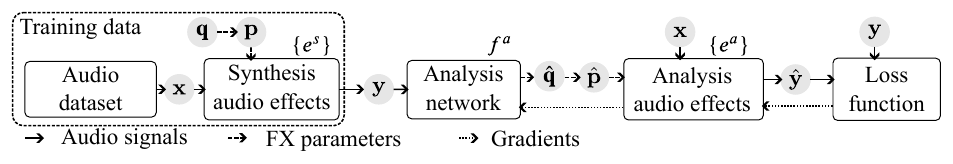}
    \caption{Proposed auto-encoder approach for \acf{beafx}.}
    \label{fig:analysis_net}
\end{figure*}

\vspace{-0.3cm}
\subsection{Proposal}
To solve the \ac{beafx} problem, we propose an auto-encoder approach which is illustrated in Figure~\ref{fig:analysis_net}.
In the left part, we construct synthetic processed mixes by applying a set of (synthesis) audio effects $\{e^s\}$ with known parameters $\p$ to an unprocessed mix $\x$. 
The results are our ground-truth processed mixes $\y$. 
Using only $\y$, an analysis network $f^a$ then estimates the set of parameters \hp{} to be used to process \x{} with (analysis) audio effects $\{e^a\}$ to produce an estimated audio sample \hy.
The analysis network $f^a$ is trained to minimize an audio loss function between \hy{} and \y{} so that $\hy\approx\y$. 
It therefore closely matches the formulation of an auto-encoder.

%
Doing so, when $\{e^a\}$=$\{e^s\}$, $f^a$ implicitly learns to replicate the parameters $\p$ given only $\y$.
When $\{e^a\}\neq\{e^s\}$, $f^a$ learns parameters to be used for $\{e^a\}$ such that the effect of the analysis chain sounds similar to the synthesis chain.

\vspace{-0.2cm}
\subsection{Paper organization}
To be able to estimate $\p$ using this auto-encoder, we should be able to ``differentiate'' the audio effects $\{e^a\}$ (i.e. compute the derivative of their outputs $\hy$ w.r.t. their inputs $\hp$). 
We therefore discuss various implementations of the audio effects in part~\ref{sec:audio_effects}.
To be able to estimate $\p$ we should define an architecture for the analysis network $f^a$. 
We therefore discuss various possible architectures in part~\ref{sec:architecture}.
During evaluation (part \ref{sec:evaluation}), we first decide for each type of effect what is the best implementation $e^a$ (among those of \ref{sec:audio_effects}) and architecture of $f^a$ (among those of \ref{sec:architecture}) to reconstruct $\y$.
We then compare our proposed approach (audio reconstruction $\hy \approx \y$) to the previously proposed approach (parameter reconstruction $\hp \approx \p$).
Finally, we evaluate the joint estimation of the whole \ac{afx} chain defined as the succession of an equalizer, a \ac{drc}, and a clipper.
We conclude in part~\ref{sec:conclusion} and propose future directions.

We provide the code of this study to ensure replicability as well as audio examples.\footnote{\href{https://peladeaucome.github.io/ICASSP-2024-BEAFX-using-DDSP/}{https://peladeaucome.github.io/ICASSP-2024-BEAFX-using-DDSP/}}

\vspace{-0.4cm}
\section{Proposal}

\subsection{Audio effects}
\label{sec:audio_effects}

In this work, we only consider the 3 following \acp{afx} commonly used for mastering~\cite{zolzerDAFXDigitalAudio2011}: 
- an \textbf{equalizer} (a cascade of linear filters which modifies the level of different frequency bands), 
- a \textbf{\acl{drc}} (which reduces the signal level when it is too loud, leaving it untouched otherwise), and 
- a \textbf{soft clipper} (which clips the signal peaks producing harmonic distortion).
We distinguish their implementation for synthesis $\{e^s\}$ and for analysis $\{e^a\}$.

$\{e^s\}$ is the implementation of the effects that have been used to create the observed master $\y$. In a real scenario, it is unknown.

$\{e^a\}$ is the implementation we use in our model to
(1) predict the parameters $\hp$ of the effects or 
(2) replicate the resulting process of the mastering. 
To perform (1) (comparing the estimated $\hp$ to the ground-truth $\p$) the implementation of the effect in $\{e^s\}$ and $\{e^a\}$ should be similar.
To perform (1) and (2), the implementation in $\{e^a\}$ should be differentiable (to estimate the parameters) or, if not, we will have to use neural networks to approximate the effects.

We now detail the implementation of the three \acp{afx} for synthesis and analysis and list them in Table~\ref{tab:summary:effect}.
The audio is normalized to 0\,dBFS before each effect

\begin{table}[h]
    \centering
    \caption{Considered audio effects and their implementations.}
    \label{tab:summary:effect}
    {\small
    \begin{tabular}{l|l|c|c}
        \toprule
        Effect            & Implementation & Synthesis & Analysis \\
        \midrule
        \textbf{Equalizer}   & Parametric  & $\surd$         & $\surd$ \\
        & Graphic     &          & $\surd$ \\
        \midrule
        \textbf{Compressor} & DSP  & $\surd$ & \\
        & Simplified DSP  & & $\surd$ \\
        & \acs{np}  & & $\surd$ \\
        & Hybdrid \acs{np}  & & $\surd$ \\
        \midrule
        \textbf{Clipper} & Parametric  &  $\surd$ & $\surd$\\
        & Taylor    &  &  $\surd$ \\
        & Chebyshev &  &  $\surd$ \\
        \bottomrule
    \end{tabular}
    }
\end{table}

\subsubsection{Equalizer}
For \textbf{synthesis}, we use a 5-band parametric equalizer: 1 low-shelf, 3 peak, 1 high-shelf. 

For \textbf{analysis}, we use either this parametric equalizer or a 10-band graphic equalizer.
Each band of the graphic equalizer has a bandwidth of 2 octaves \cite{bristow-johnsonRBJAudioEQcookbook2005}.

Each parametric band has 3 parameters: center frequency, gain, and quality factor.
Each band of the graphic equalizer has only 1 parameter: its gain.

Frequencies of parametric bands are logarithmic parameters, gains (in dB) and quality factors are linear.
Differentiable filters are implemented in the frequency domain~\cite{nercessianNeuralParametricEqualizer2020} as we find the time-aliasing error small enough for training a neural network.

\vspace{-0.3cm}
\subsubsection{Dynamic range compressor (DRC)}

\textbf{For synthesis}, we use the DSP compressor proposed in~\cite{giannoulisDigitalDynamicRange2012}.
It has 5 parameters: threshold, ratio, attack time, release time, and knee width.

\textbf{For analysis}, we either use a simplified DSP compressor or a \ac{np}. 
The simplified DSP compressor is the DSP compressor of \cite{giannoulisDigitalDynamicRange2012} but with the attack and release time linked, as proposed by~\cite{steinmetzStyleTransferAudio2022} to reduce the computation time.
The \ac{np} compressor \cite{steinmetzEfficientNeuralNetworks2022} is trained to approximate a DSP compressor%
\footnote{
To train it, we first process a set of $\x$ with the DSP compressor using known parameters $\p$. The output provides ground-truths $\y$.  
We then train the parameters $\theta$ of the \ac{np} compressor conditioned with the same parameters $\p$ such that its output $\hy = f_{\theta}(\x;\p) \approx \y$.}.
It uses a \ac{tcn} conditioned with \acs{film} \cite{perezFiLMVisualReasoning2018} layers.
In \cite{steinmetzEfficientNeuralNetworks2022}  the \ac{np} directly outputs $\hy$.
In our case, we use the same \ac{tcn} architecture but propose to replace its output activation with a sigmoid such that it provides the compressor gain factor $\mathbf{g}$ to be applied over time $n$: $\hat{y}[n]=g[n]\cdot x[n]$.~\footnote{
    We found by experiment that this modification allows to largely reduce the number of parameters (number of \ac{tcn} channels) with equivalent performances.
    With 8 channels, our causal model with a receptive field of 3\,s duration has a test mean average error (MAE) of 0.0060 while the \ac{tcn} from \cite{steinmetzEfficientNeuralNetworks2022} has a test MAE of 0.050.
}

Once trained, the \ac{np} compressor, being differentiable, can be inserted in the analysis chain $\{e^a\}$ to train the analysis network $f^a$ and obtain its compressor parameters $\hp$.
We can of course use $\hp$ to process $\x$ with the \ac{np} compressor 
but also use $\hp$ to process $\x$ with the DSP compressor.
We name the latter ``hybrid \ac{np} compressor''. 
Since it is not differentiable, we only use it during validation and testing (not during training).
It has already been used in~\cite{steinmetzStyleTransferAudio2022}.


The compressors' ratio and time parameters are logarithmic while their threshold and knee (both in dB) are linear.
%
\subsubsection{Clipper}

We propose 3 implementations of the clipper: parametric (both for synthesis and analysis), Taylor, and Chebyshev (only for analysis).

The \textbf{parametric} clipper is defined by the function $f$ defined in eq. (\ref{eq:clipper:param}) with hardness parameter $h$ blending between tanh, cubic, and hard clipping:
\begin{customequation}
    \label{eq:clipper:param}
    f(x,h)=
    \begin{cases}
        (1-h) \tanh(x) + h f_\text{cubic}(x), & h\in[0;1],
        \\
        (2-h) f_\text{cubic}(x) + (h-1) f_\text{hard}(x), & h\in(1;2].
    \end{cases}
\end{customequation}
with :
\begin{equation}
    \begin{aligned}
        &f_\text{hard}(x)=\max(-1, \min(1, x))\\
        &f_\text{cubic}(x)=
        \begin{cases}
            x+4x^3/27, &x\in[-\frac{2}{3}; \frac{2}{3}],\\
            \sgn(x),  &|x|>\frac{2}{3}.
        \end{cases}
    \end{aligned}
\end{equation}
The effect used for synthesis is constructed with the following parameters: gain $g$ (in dB), offset $o$, and hardness $h$.
\begin{customequation}
    y[n] =  \left( f(g \cdot x[n] + o, h) - f(o, h) \right)/g.
\end{customequation}

We also use two other memory-less models. 
Both have been proposed for memory-less distortion identification. 
The \textbf{Taylor} clipper is inspired by Taylor series expansions~\cite{avilaMLEstimationMemoryless2014}:
\begin{customequation}
    y[n] = \sum_{h=0}^{H-1}g_h x[n]^h.
\end{customequation}

The \textbf{Chebyshev} clipper is inspired by Chebyshev's polynomials. It has been used for non-linear audio effect identification~\cite{novakChebyshevModelSynchronized2010}:
\begin{customequation}
    y[n] = \sum_{h=0}^{H-1} g_h T_h(y[n]).
\end{customequation}
with $g_h\in[-1;1]$ being the effect's parameters and $T_n$:
\begin{equation}
\begin{aligned}
     &T_n(x) = 2xT_{n-1}(x)-T_{n-2}(x),\qquad \forall n\geq2,
     \\
     &T_0(x)=1,\qquad T_1(x)=x.
\end{aligned}
\end{equation}
In both cases, the parameters to be estimated are the $\{g_h\}$ and we set $H=24$.
All clipper parameters are linear.

\subsubsection{Parameter ranges}

All \ac{afx} parameters, $p_c / \hat{p}_c \in [p_{c, \text{min}}; p_{c, \text{max}}]$, are derived from ``normalized'' \ac{afx} parameters, $q_c / \hat{q}_c$ $\in [0;1]$.
Linear parameters (see above) are derived from $q_c / \hat{q}_c$ using an affine transformation:
\begin{customequation}
    p_c = (p_{c, \text{max}} -p_{c, \text{min}}) q_c +p_{c, \text{min}}.
\end{customequation}

Logarithmic parameters (see above) are derived from $q_c / \hat{q}_c$ using an exponential transformation:
\begin{customequation}
    p_c
    =
    e^{\left(\log\left(p_{c,\text{max}}\right) - \log\left(p_{c,\text{min}}\right)\right)q_c}p_{c,\text{min}}.
\end{customequation}

\subsection{Analysis network $f^a$}
\label{sec:architecture}

The analysis network is divided into two parts:
\begin{enumerate}
\item An \textbf{encoder}; which outputs a time invariant embedding. 
We describe and compare below 3 architectures for this encoder.
\item A MLP with 4 layers of size 2048, 1024,  512, and $C$ where $C$ is the total number of parameters $\hp$ of the effect chain $\{e^a\}$.
Each hidden layer is followed by a Batchnorm-1D and a PReLU. The output layer, which estimates normalized parameters \hq, is followed by a sigmoid.
\end{enumerate}

For the encoder, we compare three popular architectures commonly used in the Music Information Retrieval field:
\begin{description}
    \item[MEE] the Music Effects Encoder proposed by \cite{kooEndtoendMusicRemastering2022}. It consists of a cascade of residual 1D convolutional layers,
    \item[TE] a Timbre Encoder inspired by \cite{ponsTimbreAnalysisMusic2017}. It consists of a single 2D convolution layer with multiple sizes of filters. The conv-2D is applied on the CQT \cite{brownCalculationConstantSpectral1991} of the signal as implemented in \cite{cheukNnAudioOntheflyGPU2020},
    \item[TFE] a Time+Frequency Encoder inspired by \cite{ponsDeepNeuralNetworks2019}. It consists of two 2D convolutional nets, one focusing on highlighting temporal motifs, and the second on frequency motifs. The network's input is the CQT of the signal.
\end{description}

MEE has 88M parameters, TE 2.8M, and TFE 3.4M.

\section{Evaluation}
\label{sec:evaluation}



\subsection{Dataset}

For evaluation, we use the \textit{mix files} of the MUSDB18~\cite{rafiiMUSDB18CorpusMusic2017} dataset.
From those, we extract randomly clips of 10\,s duration.
Those are then converted to mono and peak-normalized to 0\,dBFS.
For each, we randomly pick the normalized parameters $\mathbf{q}\sim\mathcal{U}(0, 1)$, convert them to $\mathbf{p}$ and apply them to the clip.
We use the training, validation, and testing splits proposed by MUSDB18~\cite{rafiiMUSDB18CorpusMusic2017}.

\subsection{Training}

In the following, we compare 2 approaches for training $f^a$:
\begin{description}
    \item[Audio reconstruction $\hy \approx \y$] (our proposal): we minimize the $\ell^1$ norm between the log-magnitude Mel-spectrograms  of $\mathbf{y}$ and $\hatbf{y}$ as implemented in \cite{steinmetzAuralossAudiofocusedLoss2020}
    \footnote{We consider this metric as a surrogate for ``audio quality'' and are aware that it does not cover the whole extent of audio quality. Note that this loss has been previously used, for example in \cite{suBandwidthExtensionAll2021}.}
:
\begin{customequation}
    \Lyy =
    \norm{
        \log \left\{ \text{Mel}(|\hatbf{y}| )\right\}
        -
        \log \left\{ \text{Mel}(|\mathbf{y}|) \right\}
    }_1.
\end{customequation}
\item[Parameter reconstruction $\hq \approx \q$] (previously proposed approach): we minimize the $\ell^2$ norm between \hq{} and \q:
\begin{customequation}\label{eq:loss:params}
    \Mqq{} = \frac{1}{C}\sum_{c=0}^{C-1}(\hat{q}_c - q_c)^2.
\end{customequation}
\end{description}

\textit{Training details:} Each model $f^a$ is trained using the ADAM algorithm with a learning rate of $10^{-4}$ and a batch size of 16 during a maximum of 400 epochs, where a single epoch is defined as 430 training examples (5 from each song of the training subset).
The learning rate is scheduled to decrease by a factor of 10 when the best validation score has not been improved for 30 epochs. Training stops after 150 epochs without improvement.
To ensure the reliability of computed scores the validation is run 5 times and the test 10 times.

\subsection{Performance metrics.}
For evaluation, we compute the following metrics:
\begin{itemize}
    \item \Myy: the MSE between $\hatbf{y}$ and $\mathbf{y}$
    \item \Lyy as defined above
    \item \Mqq: the MSE between $\hatbf{q}$ and $\mathbf{q}$
\end{itemize}

To make audio loss and metrics invariant to sound level, we normalize \y{} and \hy{} by their respective RMS values\footnote{Note that this normalization is applied as a static gain on the whole signal and therefore does not interfere with the dynamic gain of the \ac{drc}.}.

\subsection{Results}
\label{sec:evaluation:single}

\subsubsection{Single effect estimation.}

For each type of effect we first decide what is the best implementation (among those of \ref{sec:audio_effects}) and architecture (among those of \ref{sec:architecture}) to reconstruct $\y$.
For each, we also indicate: 
- \textbf{Random $\hq$}: the results obtained with a random choice of $\hq$ (rather than the estimated one),
- \textbf{$\mathcal{L}(\x, \y)$}: the value of the loss when comparing the input $\x$ to the output $\y$.

For the \textbf{equalizer} (Table~\ref{tab:estimation:eq}), Parametric and Graphic provide similar results.
Since \Lyy{} indicates the difference between spectras, it is more suited to measure the performances of an EQ than \Myy.
We therefore focus on \Lyy{} and conclude that the best (0.32) configuration is to use the Parametric implementation for $\{e^a\}$ and TFE for $f^a$.

For the \textbf{compressor} (Table~\ref{tab:estimation:comp}), the best (0.011, 0.076) configuration is to use the Hybrid \ac{np} for $\{e^a\}$ and use MEE for $f^a$. 
As a reminder, the Hybrid \ac{np} compressor uses the \ac{np} compressor to estimate $\hp$ but the DSP compressor (the same used for synthesis) to get $\hy$.
This works better than using the \ac{np} compressor (0.014, 0.098) or the simplified DSP compressor (0.041, 0.16)
This is because the latter links the attack and release time parameters which might be too restrictive.
The fact that the Hybrid \ac{np} works better indicates that our proxy performs well enough for the task of estimating parameters usable with the DSP Compressor.

For the \textbf{clipping} (Table~\ref{tab:estimation:clip}), the best configuration is to use the Parametric clipper  (the same used for synthesis) for $\{e^a\}$ and use MEE for $f^a$. 

It should be noted that in all three cases, the best results are obtained when $\{e^a\}=\{e^s\}$.

\begin{table}[th!]
    \centering
    \caption{Results by implementation of $\{e^a\}$ and encoder $f^a$ for equalisation. The best results are indicated in bold. 
    }
    {\footnotesize
    \begin{tabular}{@{}l c c c c@{}}
        \toprule
        Equalizer & \multicolumn{2}{c}{Parametric} & \multicolumn{2}{c}{Graphic}\\
        \cmidrule(r){2-3}\cmidrule(r){4-5}
        Metrics
        &
        \Myy & \Lyy
        &
        \Myy & \Lyy
        \\
        \midrule
        MEE                                   & 0.35 & 0.41 & 0.38 & 0.41\\
        TE                                & 0.35 & 0.46 & 0.39 & 0.48\\
        TFE                      & 0.28 & \textbf{0.32} & \textbf{0.25} & 0.34\\
        \midrule
        Random $\hat{\mathbf{q}}$         & 0.86 & 0.89 & 1.1  & 1.1 \\
        $\mathcal{L}(\mathbf{x}, \mathbf{y})$ & 0.48 & 0.64 & 0.48 & 0.63\\
        \bottomrule
    \end{tabular}
    }
    \label{tab:estimation:eq}
    \vspace{.1cm}
    \centering
    \caption{Same for dynamic range compression. }
    {\footnotesize
    \begin{tabular}{@{}l p{.9cm} p{.8cm} p{.9cm} p{.8cm} p{.9cm} p{.8cm}@{}}
        \toprule
        Comp. & \multicolumn{2}{c}{\ac{np}} & \multicolumn{2}{c}{Hybrid \ac{np}} & \multicolumn{2}{c}{Simpl. DSP}\\
        \cmidrule(r){2-3}\cmidrule(r){4-5}\cmidrule(r){6-7}
        Metrics
        &
        \Myy & \Lyy
        &
        \Myy & \Lyy
        &
        \Myy & \Lyy
        \\
        \midrule
        MEE                  & 0.014 & 0.098 & \textbf{0.011} & \textbf{0.076} & 0.041 & 0.16\\
        TE                   & 0.020 & 0.12  & 0.019 & 0.11  & 0.042 & 0.17\\
        TFE                  & 0.015 & 0.11  & 0.012 & 0.086 & 0.042 & 0.17\\
        \midrule
        Random $\hat{\mathbf{q}}$         & 0.038 & 0.16  & 0.036 & 0.15  & 0.13  & 0.66\\
        $\mathcal{L}(\x, \y)$ & 0.041 & 0.16  & 0.041 & 0.16  & 0.040 & 0.16\\
        \bottomrule
    \end{tabular}
    }
    \label{tab:estimation:comp}
    \vspace{.1cm}
    \centering
    \caption{Same for clipping}
    {\footnotesize
    \begin{tabular}{@{}l p{.9cm} p{.8cm} p{.9cm} p{.8cm} p{.9cm} p{.8cm}@{}}
        \toprule
        Clipper & \multicolumn{2}{c}{Parametric} & \multicolumn{2}{c}{Taylor} & \multicolumn{2}{c}{Chebyshev}\\
        \cmidrule(r){2-3}\cmidrule(r){4-5}\cmidrule(r){6-7}
        Metrics
        &
        \Myy & \Lyy
        &
        \Myy & \Lyy
        &
        \Myy & \Lyy
        \\
        \midrule
        MEE                & $\mathbf{0.0045}$ & $\mathbf{0.072}$ & 3.4 & 0.15 & 0.65 & 0.17\\
        TE                 & 0.011     & 0.090 & 2.9 & 0.21 & 3.2  & 0.15\\
        TFE                & $0.0067$ & 0.075 & 2.1 & 0.16 & 1.34  & 0.12\\
        \midrule
        Random \hq         & 0.078 & 0.29 & 1.9 & 0.39 & 2.0 & 1.1\\
        $\mathcal{L}(\mathbf{x}, \mathbf{y})$ & 0.078 & 0.34 & 0.079 & 0.34 & 0.079 & 0.34\\
        \bottomrule
    \end{tabular}
    }
    \label{tab:estimation:clip}
    \vspace{.1cm}
    \centering
    \caption{Comparison between training using \Lyy{} and \Mqq{} for single effects and whole FX chain.}
    {\footnotesize    
\begin{tabular}{l p{.8cm} p{.6cm} p{.8cm} p{.8cm} p{.6cm} p{.8cm}}
        \toprule
        Loss & \multicolumn{3}{c}{\Lyy} & \multicolumn{3}{c}{\Mqq}\\
        \cmidrule(r){2-4}\cmidrule(r){5-7}
        Metrics & \Myy & \Lyy & \Mqq{} & \Myy & \Lyy & \Mqq{} \\
        \midrule
        Eq. & 0.28 & \textbf{0.32} & 0.089 & \textbf{0.27} & 0.40 & \textbf{0.072}\\
        Comp. & 0.011 & 0.076 & 0.11 & $\mathbf{0.0081}$ & \textbf{0.069} & \textbf{0.069}\\
        Clip.  & 0.0045 & 0.072 & 0.044 & $\mathbf{0.0041}$ & \textbf{0.064} & \textbf{0.028}\\
        \midrule
        Chain    & \textbf{0.31} & \textbf{0.40} & 0.10 & 0.33 & 0.49 & \textbf{0.072}\\
        \bottomrule
    \end{tabular}
    }
    \label{tab:estimation:params}
\end{table}

\subsubsection{Training method comparison.} 
We now compare our proposed training method (based on audio reconstruction $\hy \approx \y$) to the previously proposed one (based on parameter reconstruction $\hq \approx \q$).
Results are indicated in Table~\ref{tab:estimation:params}.
For each single effect, we use the best configuration found above: $f^a$=TFE for equalizer, MEE for compression and clipping. 

For \textbf{equalization}, in terms of audio quality (\Lyy), the network trained to minimize \Lyy{} outperforms (0.32) the one that minimizes \Mqq{} (0.40). 
But in terms of parameter estimation (\Mqq) minimizing directly \Mqq{} leads to better results (0.072).

For \textbf{compression} and \textbf{clipping}, training by minimizing the parameter distance (\Mqq) leads to better results both in terms of audio quality (\Lyy=0.069, 0.064) and parameter estimation (\Mqq=0.069, 0.028).

\subsubsection{Effects chain estimation} 
We finally evaluate the estimation of the whole \textbf{chain} of effect (equalizer$\to$compressor$\to$clipper).
In this case, we use $f^a$=TFE for all.
As for audio quality, we see (row ``Chain'' in Table~\ref{tab:estimation:params}) that our approach (minimizing \Lyy) leads to the best results: \Myy=0.31 and \Lyy=0.40.
However, as can be predicted, minimizing directly $\Mqq$ leads to better estimation of the parameters: \Mqq=0.072.
These contrasting outcomes underscore that achieving accurate parameter estimation ($\hp \approx \p$) does not guarantee high audio quality ($\hy \approx \y$).
While our approach may not yield the best parameter estimation, it does yield the best audio transform estimation.

\section{Conclusion}
\label{sec:conclusion}

In this work, we proposed an auto-encoder approach for \acl{beafx}.
Given only processed (wet) audio signals, we train a neural network to estimate \ac{afx} parameters such that when used for effects applied to an unprocessed (dry) signal it approximates the processed (wet) signal.
This allows training a network using real dry/wet data pairs without knowing the exact effect implementation.
We show that our audio-based method better replicates the audio quality of the mastering process than the previous parameter-based method.

Further work will focus on performing subjective perceptual experiments, including other important mastering effects in the chain and testing their estimation on real mastered music productions.


\end{document}